\documentstyle[12pt,psfig]{article}

\begin{document}
\begin{flushright}
SLAC-PUB-7968\\
October 1998
\end{flushright}
\vfill
\begin{center}
{\Large {High Density Parton Shadowing Corrections in DIS Scaling Violations}
\footnote{\baselineskip=13pt Work supported in part by the Department
of Energy, contract DE--AC03--76SF00515, by CNPq,  and by 
Programa de Apoio a N\'ucleos de Excel\^encia (PRONEX), BRASIL.
}}

\vspace{15mm}
{\bf M. B. Gay Ducati}\\
\vspace{5mm}
{\em Instituto de F\'{\i}sica, Universidade Federal do Rio Grande do Sul\\ 
Caixa Postal 15051, CEP 91501-970, Porto Alegre, RS, BRASIL\\
E-mail: gay@if.ufrgs.br} \\
and \\
{\em Stanford Linear Accelerator Center \\
Stanford University, Stanford, California 94309}\\

\vspace{10mm}
{\bf Victor P.  Gon\c{c}alves} \\ 
{\em Instituto de F\'{\i}sica, Universidade Federal do Rio Grande do Sul\\ 
Caixa Postal 15051, CEP 91501-970, Porto Alegre, RS, BRASIL \\
E-mail:barros@if.ufrgs.br}
\end{center}
\bigskip
\begin{center}
Submitted to Physical Review D. 
\end{center}

\vfill

\newpage

\begin{center}
Abstract
\end{center}

The description of the dynamics at high density parton regime is  one of the  main open 
questions of the strong interactions theory. In this paper we address the 
shadowing corrections (SC) in the scaling violations of the $F_2$ structure function 
using the eikonal approach. We propose a procedure to estimate the distinct contributions 
to  the SC for $F_2$ and its slope and  show that the recent ZEUS data can be described 
if the SC in the quark and  gluon sectors are considered. The radius dependence of the  
SC is estimated.  Moreover, we calculate  the   
superior limit above  which the unitarity 
corrections cannot be disregarded at low $Q^2$ and show that the recent 
HERA data overcomes   this bound.

\bigskip
\begin{center}
\end{center}

\newpage

\section{Introduction}

The description of the dynamics at high density parton regime is one of the  main open 
questions of the strong interactions theory. While in the region of moderate Bjorken $x$ 
($x \ge 10^{-2}$) the  well-established methods of 
operator product expansion and renormalization group equations  have been applied successfully, 
the small $x$ region still lacks a consistent theoretical framework (For a review see \cite{cooper}). 
Basically, its is questionable the use of  
 the DGLAP equations \cite{dglap}, which reflects the dynamics at moderate $x$, in 
the region of small values of $x$. The traditional procedure of using the DGLAP equations  to 
calculate the gluon distribution at small $x$ and large momentum transfer $Q^2$ is by summing 
the leading powers of  $\alpha_s\,ln\,Q^2\,ln(\frac{1}{x})$, where $\alpha_s$ is  the strong coupling constant, known as   the double-leading-logarithm 
approximation (DLLA). In axial gauges, these leading double logarithms are generated by 
ladder diagrams in which the emitted gluons have strongly ordered transverse
momenta, as well as strongly ordered longitudinal momenta.
Therefore the DGLAP must breakdown at small values of $x$, firstly because this 
framework does not account for the  contributions to the cross section 
which are leading  in $\alpha_s \, ln (\frac{1}{x})$ \cite{bfkl}. Secondly, because the 
parton densities  become large and there is  need to develop a high density 
formulation of QCD \cite{100}.

There has been intense debate on to which extent non-conventional QCD evolution is 
required by the deep inelastic $ep$ HERA data \cite{cooper,hera96}. Good fits to the $F_2$ data 
for $Q^2 \ge 1\,GeV^2$ can be obtained from 
distinct approaches, which consider DGLAP and/or BFKL evolution equations 
\cite{ball,martin}.  In particular, the conventional  
perturbative QCD approach is very successful in describing the main features 
of HERA data and, hence, the signal of non-conventional QCD dynamics is 
hidden or mimicked by a strong background of conventional QCD evolution.
Our goal in this paper is the role of  the  shadowing corrections (SC) in $F_2$ and its slope. In the last twenty years, 
several authors (see  \cite{sha} for some phenomenological analysis) have performed a detailed study of the shadowing 
effect although without a strong 
experimental evidence of this effect in the data, mainly since the main observable, 
the $F_2$ structure function, is inclusive to the effects in the gluon distribution.  Recently 
we   have estimated the shadowing corrections to the  $F_2^c$ and $F_L$ at  
HERA kinematic region using the eikonal 
approach \cite{prd}. These observables are directly dependent on the behavior 
of the gluon distribution. We  have shown that the shadowing corrections to these 
observables are important, however  the experimental  errors in these 
observables are still large to allow a  discrimination between our predictions 
and the DGLAP predictions.
Here   we estimate the shadowing corrections to the scaling violations 
of the proton structure function. Basically, there are two possibilities to estimate 
the SC using the eikonal approach. We can calculate  damping factors, which 
represent the ratio between the observable with and without shadowing, and  subsequently  
apply these factors in the conventional DGLAP predictions. This procedure was 
used in refs. \cite{glmn1,glmn}, also considering a two radius model for the nucleon. In this paper we propose a second  procedure to estimate the SC in DIS, 
where  the observables are directly calculated in the eikonal approach and the distinct 
contributions to the SC are analysed in the same approach, reducing the number 
of free parameters. A larger discussion about the distinct procedures is made in section II.  

The recent HERA data on the slope of the 
$F_2$ structure function  \cite{zeus} present at small values of $x$ and $Q^2$ a different 
behavior than predicted  by the standard DGLAP framework. Basically, the HERA 
data present a `turn over' of the slope around $x \approx 10^{-4}$, which cannot be 
described using the GRV94 parametrization \cite{grv95} and the DGLAP evolution 
equations. We show that this behavior is predicted 
by the eikonal approach considering the shadowing corrections for the gluon and quark sectors.

The value of the shadowing corrections depends crucially on the size of the target $R$. 
The value of the effective radius  $R$ depends on how the gluon ladders couple to 
the proton; {\it i.e.}, on 
how the gluons are distributed within the proton \cite{hotspot}. In this paper we estimate 
the $R$ dependence of the SC. We show that the HERA data on the $F_2$ 
and its slope  can be described consistently using $R^2 = 5\,GeV^{-2}$. This   value  
agrees with the HERA results on the diffractive  $J/\Psi$ photoproduction \cite{zeusjpsi,h1jpsi}.

 The steep increase of the gluon distribution predicted by DGLAP and BFKL equations at 
high energies would 
eventually violate the Froissart bound \cite{froi}, which restricts the rate of 
growth of the total cross section to $ln^2(\frac{1}{x})$. This bound may not be 
applicable in the case of particles off-mass shell \cite{yndu}, but in this paper we present an 
approach for this problem. Basically, we estimate a  limit below which the unitarity 
corrections may be disregarded and  show that the recent HERA data surpass this 
 boundary, as  predicted in \cite{plb}, at small values of $x$ and $Q^2$.

This paper is organized as follows. In section II, the eikonal approach and the shadowing 
corrections for $F_2$ and its slope are considered. We estimate the distinct contributions 
for the SC and demonstrate that the $\frac{dF_2(x,Q^2)}{dlogQ^2} $ data may be 
described considering the shadowing in the gluon and quark sectors. In section III, we 
estimate the $R$ dependence of the shadowing corrections. In section IV, we present 
a boundary related to unitarity for $F_2$ and $\frac{dF_2(x,Q^2)}{dlogQ^2} $ and show 
that the actual HERA data for small $x$  and $Q^2$ overcomes this boundary. Therefore, 
the shadowing corrections should be considered in the calculation of the  observables  in 
this kinematic region.
Finally, in section V, we present a summary of our results.

\section{The Shadowing Corrections in pQCD}

  The deep inelastic scattering (DIS) is usually described in a frame where the proton is 
going very fast. In this case the  shadowing effect is a result of an overlap  of the parton 
clouds in the longitudinal direction. Other interpretation of DIS is the intuitive view  
proposed by V. N. Gribov many years ago for 
the DIS on  nuclear targets \cite{gribov}. Gribov's assumption is that at small values of $x$  the virtual 
photon fluctuates into a $q\overline{q}$ pair well before the interaction 
with the target, and this system interacts with
the target. This formalism has been established as an useful tool for 
calculating deep inelastic and related diffractive cross section for 
$\gamma^*\,p$ scattering in the last years \cite{nik,buch}.
The Gribov factorization follows from the fact that the lifetime of the $q\overline{q}$
fluctuation is much larger than the time of the interactions with partons. According to 
the uncertainty principle, the fluctuation time is $\approx \frac{1}{m\,x}$, 
where $m$ denotes the target mass.

The space-time picture of the DIS in the target
rest frame can be viewed as the decay of the virtual photon at high energy
(small $x$) into a quark-antiquark pair long before the 
interaction with the target. The $q\overline{q}$ pair subsequently interacts 
with the target.  In the small $x$ region, where 
$x \ll \frac{1}{2mR}$ ($R$ is the size of the target), the $q\overline{q}$  pair 
crosses the target with fixed
transverse distance $r_t$ between the quarks. It allows to factorize the total 
cross section between the wave function of the photon and the interaction 
cross section of the quark-antiquark pair with the target. The photon wave function 
is calculable and the interaction cross section is modeled. Therefore we have that the 
proton structure function is given by \cite{nik}
\begin{eqnarray}
F_2(x,Q^2) = \frac{Q^2}{4 \pi \alpha_{em}} \int dz \int d^2r_t |\Psi(z,r_t)|^2 \, \sigma^{q\overline{q}}(z,r_t)\,\,,
\label{f2target}
\end{eqnarray}
where 
\begin{eqnarray}
|\Psi(z,r_t)|^2 = \frac{6 \alpha_{em}}{(2 \pi)^2} \sum^{n_f}_i e_f^2 \{[z^2 
+ (1-z)^2] \epsilon^2\, K_1(\epsilon r_t)^2 + m_f^2\, K_0^2(\epsilon r_t)^2\}\,\,,
\label{wave}
\end{eqnarray}
$\alpha_{em}$ is the electromagnetic coupling constant,
$\epsilon = z(1-z)Q^2 + m_f^2$, $m_f$ is the quark mass, $n_f$ is the number 
of active flavors, $e_f^2$ is the square of the  parton charge (in units of $e$), $K_{0,1}$ 
are the modified Bessel functions and $z$ is the fraction of the photon's light-cone 
momentum carried by one of the quarks of the pair.  In the 
leading log$(1/x)$ approximation we can neglect the change of $z$ during the 
interaction and describe the cross section $\sigma^{q\overline{q}}(z,r_t^2)$ as 
a function of the variable $x$. Considering only light quarks ($i=u,\,d,\,s$)   $F_2$ 
can be expressed by \cite{plb}
\begin{eqnarray}
F_2(x,Q^2) = \frac{1}{4 \pi^3} \sum_{u,d,s}  e_f^2 \int_{\frac{1}
{Q^2}}^{\frac{1}{Q_0^2}} \frac{ d^2r_t}{r_t^4}\,\sigma^{q\overline{q}}(x,r_t) \,\,.
\label{f2sim}
\end{eqnarray}
We have introduced a cutoff in the superior limit of the integration in order  to eliminate 
the long distance (non-perturbative) contribution in our calculations. 
In this paper we assume $Q_0^2 = 0.4\,GeV^2$ as in our previous works in this subject.
 
We estimated the shadowing corrections considering the eikonal approach \cite{chou}, 
which is formulated in the impact parameter space.  Here we review the main assumptions of the eikonal approach. 
In the impact  parameter representation, the scattering amplitude $A(s,t)$, where 
$t= - q_t^2$ is the momentum transfer squared,  is given by
\begin{eqnarray}
a(s,b_t) = \frac{1}{2\pi} \int d^2q_t \, e^{-i\vec{q_t}.\vec{b_t}} A(s,t) \,\,.
\end{eqnarray}
The total cross section is written as
\begin{eqnarray}
\sigma_{tot}(s) = 2 \int d^2b_t \, Im \,a(s,b_t)\,\,,
\label{tot}
\end{eqnarray}
and the unitarity constraint stands as 
\begin{eqnarray}
2\, Im \,a(s,b_t) = |a(s,b_t)|^2 + G_{in}(s,b_t)
\label{uni}
\end{eqnarray}
at fixed $b_t$, where $G_{in}$ is the sum  of all inelastic channels. For high energies 
the general solution of   Eq. (\ref{uni}) is:
\begin{eqnarray}
a(s,b_t) = i \left[1 - e^{-\frac{\Omega(s,b_t)}{2}}\right]\,\,,
\label{soluni}
\end{eqnarray}
where the opacity $\Omega(s,b_t)$ is a real arbitrary function, which is modeled in the 
eikonal approach. 

Using the $s$-channel unitarity constraint (\ref{soluni}) in the expression (\ref{f2sim}),  
the $F_2$ structure function can be written in the eikonal approach as \cite{ayala2}
\begin{eqnarray}
F_2(x,Q^2) =  \frac{1}{2\pi^3} \sum_{u,d,s} e_f^2 \int_{\frac{1}{Q^2}}^{\frac{1}{Q_0^2}} \frac{d^2r_t}{r_t^4} \int d^2b_t 
\{1 - e^{-\frac{1}{2}\Omega_{q\overline{q}}(x,r_t,b_t)}\}\,\,,
\label{f2eik}
\end{eqnarray}
where the opacity $\Omega_{q\overline{q}}(x,r_t,b_t)$ describes the interaction 
of the $q\overline{q}$ pair with the target.

In the region where  $\Omega_{q\overline{q}}$ is small $(\Omega_{q\overline{q}} \ll 1)$ the  
$b_t$ dependence can be factorized as $\Omega_{q\overline{q}} = \overline{\Omega_{q\overline{q}}} S(b_t)$ \cite{100}, 
with the normalization $\int d^2b_t\, S(b_t) = 1$.  The eikonal approach assumes that 
the factorization of the $b_t$ dependence 
$\Omega_{q\overline{q}} = \overline{\Omega_{q\overline{q}}} S(b_t)$, which is 
valid in  the region where  $\Omega_{q\overline{q}}$ is small, occurs in the whole kinematical region.
The main assumption of the eikonal approach in pQCD is the identification of opacity
$\overline{\Omega_{q\overline{q}}}$ with the gluon distribution.
In \cite{plb} the opacity  is given by
\begin{eqnarray}
 \overline{\Omega_{q\overline{q}}} = \frac{ \alpha_s}{3}\,\pi^2\,r_t^2\,
 xG(x, Q^2)\,\,,
\label{omega}
\end{eqnarray}
where    $xG(x,Q^2)$ is the gluon 
distribution.  Therefore the behavior of the $F_2$ structure function
 (\ref{f2eik}) in the small-$x$ region is mainly determined by the behavior of 
the gluon distribution in this region.

The use of the Gaussian parametrization for
the  nucleon profile function $S(b_t) = \frac{1}{\pi R^2} e^{-\frac{b^2}{R^2}}$, 
where $R$ is a free parameter,   simplifies the calculations.
In general this parameter is identified with the proton radius. However, $R$ is associated 
with the spatial gluon distribution within  the proton, which may be smaller than the 
proton radius (see discussion in the next section).

 Using the expression (\ref{omega}) in (\ref{f2eik}) and
doing the integral over $b_t$,  the  master equation for $F_2$ is obtained \cite{ayala2}
\begin{eqnarray}
F_2(x,Q^2) =  \frac{2R^2}{3\pi^2} \sum_{u,d,s} e_f^2 \int_{\frac{1}{Q^2}}^{\frac{1}{Q_0^2}} \frac{d^2r_t}{\pi r_t^4} \{C + ln(\kappa_q(x, r_t^2)) + E_1(\kappa_q(x, r_t^2))\}\,\,,
\label{diseik2}
\end{eqnarray}
where $C$ is the Euler constant,  $E_1$ is the exponential function,  and the function  
$\kappa_q(x, r_t^2) = \frac{ \alpha_s}{3 R^2}\,\pi\,r_t^2\,
 xG(x,\frac{1}{r_t^2})$. Expanding the  equation (\ref{diseik2})  for small $\kappa_q$, 
the first term (Born term) will correspond to the usual DGLAP equation in the small $x$ 
region, while the other terms will take into account the shadowing corrections.

The  slope of $F_2$ structure function in the eikonal approach is straightforward from the expression (\ref{diseik2}). We obtain that
\begin{eqnarray}
\frac{dF_2(x,Q^2)}{dlogQ^2} =  \frac{2R^2 Q^2}{3\pi^2} \sum_{u,d,s} e_f^2 
 \{C + ln(\kappa_q(x, r_t^2)) + E_1(\kappa_q(x, r_t^2))\}\,\,.
\label{df2eik}
\end{eqnarray}

The expressions (\ref{diseik2}) and (\ref{df2eik}) predict  the behavior of the shadowing 
corrections to $F_2$ and its slope considering the eikonal approach for the interaction 
of the $q\overline{q}$ with the target.  In this case we are calculating  the SC associated 
with the passage of the $q\overline{q}$ pair through the target. Following \cite{glmn} 
we will denote this contribution as the quark sector contribution to the SC.

The behavior of $F_2$ and its slope are associated with the behavior of the gluon 
distribution used as input in  (\ref{diseik2}) and (\ref{df2eik}). In general, it is assumed 
that the gluon distribution is described by a parametrization of the parton distributions 
(for example: GRV, MRS, CTEQ) \cite{grv95,mrs,cteq}. In this case the shadowing 
in the gluon distribution is not included explicitly. 
In a general case we must also estimate the shadowing corrections for the gluon 
distribution, {\it i.e.} in the quark and the gluon sectors. In this case we must estimate 
the SC for the gluon distribution using the eikonal approach, similarly to the $F_2$ case.  
This was made in \cite{ayala1} and here we only present the main steps of the approach.

The gluon distribution can be obtained in  the target
rest frame considering  the decay of a virtual gluon at high energy
(small $x$) into a gluon-gluon pair long before the 
interaction with the target. The $gg$ pair subsequently interacts 
with the target, with  the transverse distance $r_t$ between the gluons  assumed fixed. 
In this case the  cross section of the absorption of a gluon $g^*$ with virtuality $Q^2$ 
can be written as 
\begin{eqnarray}
\sigma^{g^* + \rm nucleon}(x,Q^2) = \int_0^1 dz \int \frac{d^2r_t}{\pi} 
 |\Psi_t^{g^*}(Q^2,r_t,x,z)|^2 \sigma^{gg+\rm nucleon}(z,r_t^2)\,\,,
\label{sec1}
\end{eqnarray}
where $z$ is the fraction of energy carried by the gluon and $\Psi_t^{g^*}$ is the 
wave function of the transverse polarized 
gluon in the virtual probe. Furthermore, $\sigma^{gg+\rm nucleon}(z,r_t^2)$ is the 
cross section of the interaction of the $gg$ pair with the  nucleon.  
Considering the $s$-channel unitarity and the eikonal model, equation  (\ref{sec1}) 
can be written as
\begin{eqnarray}
\sigma^{g^* + \rm nucleon}(x,Q^2) = \int_0^1 dz \int \frac{d^2r_t}{\pi} 
\int \frac{d^2b_t}{\pi} |\Psi_t^{g^*}(Q^2,r_t,x,z)|^2 
\,\left(1 - e^{-\frac{1}{2}  \overline{\Omega_{gg}} S(b_t)}\right)\,\,, \nonumber 
\label{diseik}
\end{eqnarray}
where the factorization of the $b_t$ dependence in  the opacity $\Omega_{gg}
(x,r_t,b_t)$ was assumed.
Using the relation $\sigma^{g^* + \rm nucleon}(x,Q^2) = \frac{4\pi^2 \alpha_s}{Q^2}xG(x,Q^2)$ 
and the expression of the wave $\Psi^{g^*}$ calculated 
in \cite{mueller,ayala1},  the 
Glauber-Mueller formula for the gluon  distribution is obtained as
\begin{eqnarray}
xG(x,Q^2) = \frac{4}{\pi^2} \int_x^1 \frac{dx^{\prime}}{x^{\prime}}
\int_{\frac{4}{Q^2}}^{\infty} \frac{d^2r_t}{\pi r_t^4} \int_0^{\infty}
\frac{d^2b_t}{\pi}\,2\,\left[1 - e^{-\frac{1}{2}\sigma_N^{gg}(x^{\prime}
,\frac{r_t^2}{4})S(b_t)}\right]\,\,,
\label{gluon}
\end{eqnarray}
where
$ \overline{\Omega_{gg}} =  \sigma_N^{gg}$ describes the interaction of the $gg$ pair with the target.
Using the  Gaussian parametrization for
the  nucleon profile function, doing the integral over $b_t$,  the  master equation 
for the gluon distribution is obtained as
\begin{eqnarray}
xG(x,Q^2) = \frac{2R^2}{\pi^2}\int_x^1 \frac{dx^{\prime}}{x^{\prime}}
\int_{\frac{1}{Q^2}}^{\frac{1}{Q_0^2}} \frac{d^2r_t}{\pi r_t^4} \{C 
+ ln(\kappa_G(x^{\prime}, r_t^2)) + E_1(\kappa_G(x^{\prime}, r_t^2))\}  \,\,,
\label{masterg}
\end{eqnarray} 
where the function  
$\kappa_G(x, r_t^2) = \frac{3 \alpha_s}{2R^2}\,\pi\,r_t^2\,
 xG(x,\frac{1}{r_t^2})$. Again, if equation (\ref{masterg}) is expanded for small $\kappa_G$, 
 the first term (Born term) will correspond to 
the usual DGLAP equation in the small $x$ region, while 
 the other terms will take into account the shadowing corrections. 
The expressions (\ref{diseik2}),  (\ref{df2eik}) and  (\ref{masterg}) are correct 
in the double leading logarithmic approximation (DLLA). As shown in \cite{ayala2}
 the DLLA does not work quite well in the accessible kinematic region ($Q^2 > 
0.4 \,GeV^2$ and $x > 10^{-6}$). Consequently, a more realistic approach must 
be considered to calculate the observables. In \cite{ayala2} the subtraction of the 
Born term and the addition of the GRV parametrization were proposed to the $F_2$ 
and $xG$ cases. In these cases we have
\begin{eqnarray}
F_2(x,Q^2) =  F_2(x,Q^2) \mbox{[Eq.  (\ref{diseik2})]} -  F_2(x,Q^2) 
\mbox{[Born]} +  F_2(x,Q^2) \mbox{[GRV]} \,\,\, ,
\label{f2}
\end{eqnarray}
and 
\begin{eqnarray}
xG(x,Q^2) =  xG(x,Q^2) \mbox{[Eq.  (\ref{masterg})]} -  xG(x,Q^2) 
\mbox{[Born]} +  xG(x,Q^2) \mbox{[GRV]} \,\,\, ,
\label{xg}
\end{eqnarray}
where the Born term is the first term  in the expansion in $\kappa_q$ and $\kappa_g$ 
of the equations  (\ref{diseik2}) and  (\ref{masterg}), respectively (see \cite{prd} 
for more details). Here we present this procedure for the $F_2$ slope. In this case
\begin{eqnarray}
\frac{dF_2(x,Q^2)}{dlogQ^2} =  \frac{dF_2(x,Q^2)}{dlogQ^2} \mbox{[Eq.  
(\ref{df2eik})]} -  \frac{dF_2(x,Q^2)}{dlogQ^2} \mbox{[Born]} +  
\frac{dF_2(x,Q^2)}{dlogQ^2} \mbox{[GRV]} \,\,\, ,
\label{df2}
\end{eqnarray}
where the Born term is the first term in the expansion in $\kappa_q$  of the 
equation (\ref{df2eik}). The last term is associated with  
 the traditional DGLAP framework, which 
at small values of $x$ predicts 
\begin{eqnarray}
\frac{dF_2(x,Q^2)}{dlogQ^2} = \frac{10 \alpha_s(Q^2)}{9 \pi} \int_0^{1-x} 
dz \, P_{qg}(z) \, \frac{x}{1-z}g\left(\frac{x}{1-z},Q^2\right)\,\,,
\label{df2glap}
\end{eqnarray}
where $\alpha_s(Q^2)$ is the  running  coupling  constant  and the splitting function 
$P_{qg}(x)$ gives the probability to find a quark with momentum fraction $x$ 
inside a gluon. This equation describes the scaling violations of the proton 
structure function in terms of the gluon distribution. We use the GRV parametrization 
as input in the expression (\ref{df2glap}).
\
In the general approach proposed in this paper we will use the solution of the 
equation (\ref{xg}) as input in the first terms of (\ref{f2}) and (\ref{df2}). 
As the expression  (\ref{xg}) estimates the gluon shadowing, the use of this distribution in 
 the expressions (\ref{f2}) and (\ref{df2}), which  consider  the contribution to SC 
associated with the passage of $q\overline{q}$ pair through the target, allows to 
estimate the SC to  both sectors (quark  + gluon) of the observables.  Our goal is 
the discrimination of the distinct contributions to the SC in $F_2$ and 
$\frac{dF_2(x,Q^2)}{dlogQ^2}$.

\begin{figure}
\centerline{\psfig{file=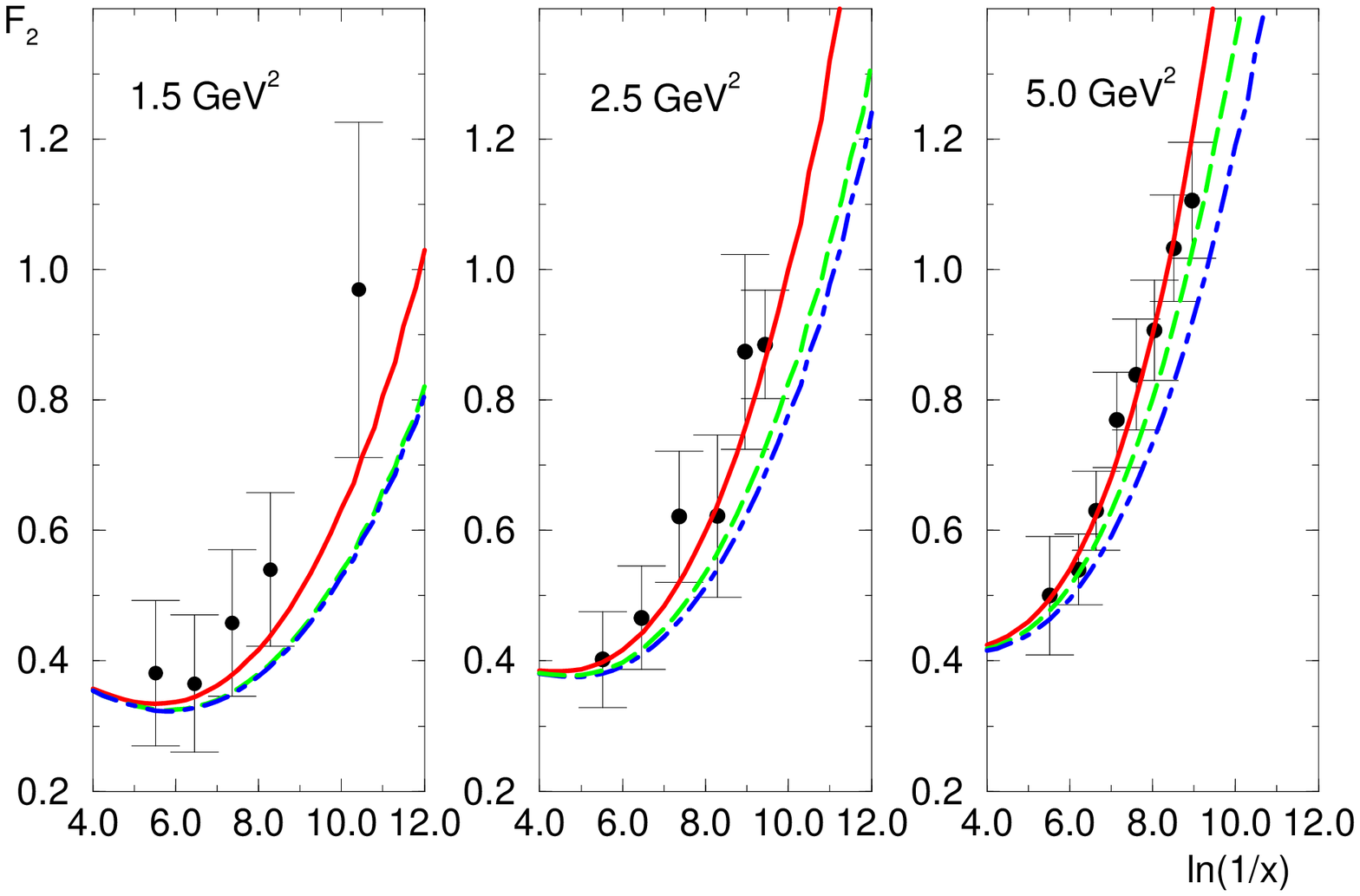,width=150mm}}
\caption{ The $F_2$ structure function as a function of the variable $ln(\frac{1}{x})$ 
for different virtualities. Data from H1 \cite{h1}.  The solid curve corresponds to GRV, the dashed curve to
SC in the quark sector, the long-short dashed curve to SC in both quark and gluon sectors.}
\label{fig1}
\end{figure}

In Fig. \ref{fig1} we present our results for the $F_2$ structure function as a function 
of the variable $ln\,(\frac{1}{x})$ for different virtualities. We have used $R^2 
= 5\,GeV^{-2}$ in these calculations. In the next section the $R$ dependence 
of our results is analysed. 
We present our results using the expression (\ref{f2}) (quark sector) and using the  
solution of the equation (\ref{xg}) as input in the first term of (\ref{f2}) (quark 
+ gluon sector). The predictions of the GRV parametrization are also shown. 
We consider the  HERA data at low $Q^2$ since for $Q^2 > 6\,GeV^2$ the 
SC start to fall down (For a discussion of the SC to $F_2$ considering the quark sector see \cite{ayala2,ayala3}). We can see that at small values of $Q^2$ the predictions 
for $F_2$ considering the quark and the quark-gluon sector are approximately 
identical. However, for larger values of $Q^2$ the predictions of the quark-gluon 
sector disagree with the  H1 data \cite{h1}. Therefore, 
the contribution of the gluon shadowing to $F_2$ in an eikonal approach 
superestimates the shadowing corrections at large $Q^2$ values.

\begin{figure}
\centerline{\psfig{file=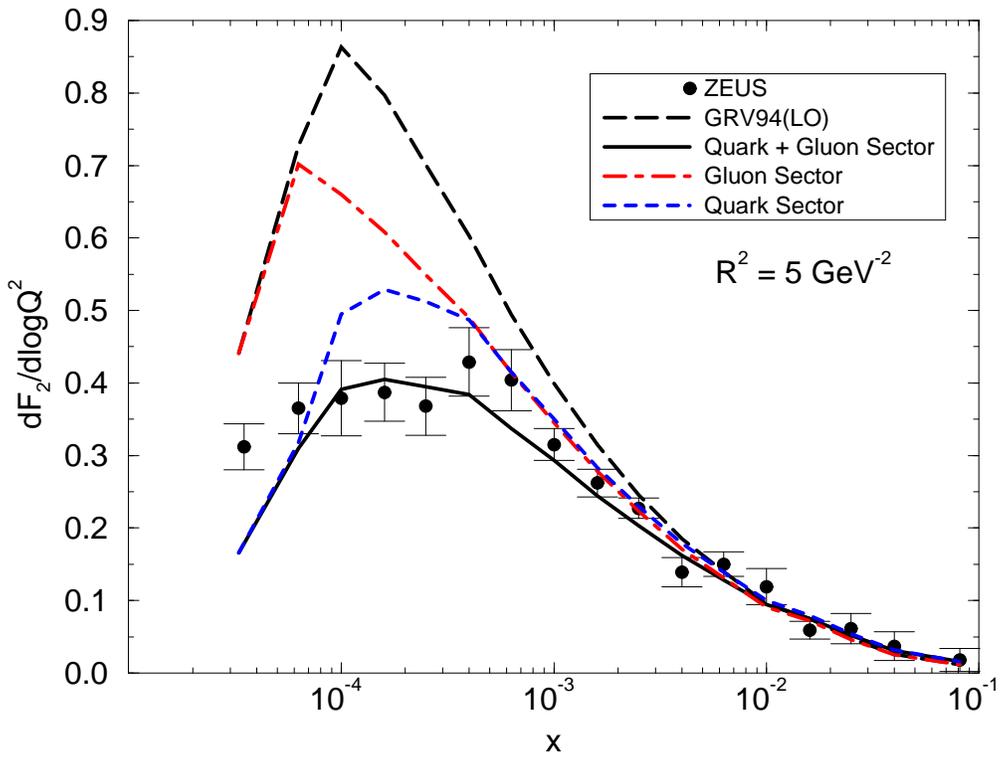,width=150mm}}
\caption{The $F_2$  slope  as a function of the variable $x$.  Data from ZEUS \cite{zeus}. The data points correspond to a different $x$ and $Q^2$ value.  The solid curve corresponds to SC in both
quark and gluon sectors.}
\label{fig2}
\end{figure}

In Fig. \ref{fig2} we present our results for the SC in the   
$\frac{dF_2(x,Q^2)}{dlogQ^2}$ as a function of $x$. 
The ZEUS data  points \cite{zeus} correspond to  different $x$ and $Q^2$ value. The 
$(x,Q^2)$ points are averaged values obtained from each of the experimental 
data distribution bins.   Only the data points with $<Q^2> \, \ge 0.52\,GeV^2$ 
and $x < 10^{-1}$ were used here.

The SC are estimated considering the expression (\ref{df2}) (quark sector) and 
using the  solution of the equation (\ref{xg}) as input in the first term of  
(\ref{df2}) (quark + gluon sector). Moreover,  the predictions of the 
traditional DGLAP framework, which 
at small values of $x$ is given by the expression (\ref{df2glap}) are also presented. 
 We can see that the DGLAP predictions fail to describe the ZEUS data at small 
values of $x$ and $Q^2$. 
However we see that in the traditional framework (DGLAP + GRV94) a 'turn over' 
is also present at small values of $x$ and $Q^2$.  Basically, this occurs since the 
smaller $Q^2$ value used ($<Q^2> = 0.52\,GeV^2$) is very near  the initial virtuality 
of the GRV parametrization, where the gluon distribution is 'valence like'. Therefore the 
gluon distribution and  the $F_2$ slope  are approximately flat in this region. For the 
second smaller value of $Q^2$ used ($<Q^2> = 1.1\,GeV^2$) the evolution length is 
larger, which implies that  the gluon distribution (and the $F_2$ slope) already presents 
a steep behavior. The link between these points implies the 'turn over' presented 
in  Fig. \ref{fig2}. The main problem is that this 'turn over' is higher than observed in the ZEUS data.
This  implies that $xG(x,Q^2)$ differs 
from the previous standard expectations in the limit of small $x$ and $Q^2$. 
This effect is not observed in the $F_2$ structure function since it is inclusive to  the 
behavior of the gluon distribution, which can be verified analysing the predictions of the 
distinct parametrizations. The gluon distribution predicted by these parametrizations 
differs in approximately 50 $\%$. 

The prediction of the gluon sector, which is 
obtained using the solution of the expression (\ref{xg}) as input in (\ref{df2glap}) is also presented.  We can see that at  
larger values of $Q^2$ and $x$ all predictions are approximately identical. However, 
at small values of $x$ and $Q^2$,   the  ZEUS data  
is not  well  described  considering only the quark or the gluon sector to the SC. The 
contribution of the gluon shadowing  is essential  in the region of small values of $x$ 
and $Q^2$, {\it i.e.} a shadowed gluon distribution should be used as input in the 
eikonalized expression (\ref{df2}) in this kinematic region.

Our conclusion is that 
at small values of $x$ and $Q^2$ it should be considered the contribution of the 
gluon shadowing to estimate the SC to $F_2$ and its slope in the eikonal approach.
While for $F_2$ the contribution of the gluon shadowing may be disregarded, it is 
essential for the $F_2$ slope.
The $\frac{dF_2(x,Q^2)}{dlogQ^2}$ data show that a consistent approach 
should consider both contributions at small $x$ and $Q^2$.

Before we conclude this section some comments are in order.
We show that the $\frac{dF_2(x,Q^2)}{dlogQ^2}$ data can be successfully 
described  considering the shadowing corrections in the quark and gluon sectors. 
A similar conclusion was obtained in \cite{glmn1,glmn}, where the eikonal approach was 
also used to estimate the SC in the quark and gluon sectors, but  a distinct procedure 
was used to estimate the SC for the $F_2$ slope. In \cite{glmn}  damping factors 
are calculated separately for both sectors and applied to the standard DGLAP predictions. 
The behavior of the gluon distribution at small values of $Q^2$ was modeled separately, since the gluon distribution (\ref{masterg}) vanish for $Q^2 = Q_0^2$.  
 This procedure introduces a   free parameter $\mu^2$, 
 beyond  the usual ones used in the eikonal approach ($Q_0^2, \, R^2$).  
The  distinct procedure
 proposed here estimates the observables  directly within the eikonal approach and the 
 shadowing corrections in the different sectors are calculated  within  the same  approach.
In our calculations   there are only two  free parameters: (i) the cutoff ($Q_0^2 = 0.4 
\, GeV^2$) in order to eliminate the long distance contribution,   and (ii) the radius 
$R$ ($R^2 = 5 \, GeV^{-2}$). The choice of these parameters is associated with 
the initial virtuality of the GRV parametrization used in our calculations, 
and the estimates 
obtained using the HERA data on diffractive photoproduction of $J/\Psi$ 
vector meson  
(see discussion in the next section) respectively \cite{zeusjpsi,h1jpsi}. 
 In our procedure the  region of 
small values of $Q^2 \approx Q_0^2$ is determined by the behavior of the 
GRV parameterization in this region, since we are using 
the  eq. (\ref{xg}) to calculate the gluon distribution.
 For $Q^2 = Q_0^2$ 
the two first terms of (\ref{xg}) vanish and the gluon distribution is described by the GRV parameterization, {\it i.e.} $xG(x,Q_0^2) = xG(x,Q_0^2) \mbox{[GRV]}$.
                                         
The eikonal approach describes the ZEUS data,  
as well as the DGLAP evolution equations using  modified parton distributions. 
Recently, the MRST group \cite{mrst} has proposed a different set of parton 
parametrizations which consider a initial 'valence-like' gluon distribution. 
This  
parametrization allows to describe the $F_2$ slope data without an unconventional effect. 
This occurs because there is  a large freedom in the initial parton distributions and the 
initial virtuality used in these parametrizations.  We believe that only   
a comprehensive  
analysis of distinct observables ($F_L, \, F_2^c, \, \frac{dF_2(x,Q^2)}{dlogQ^2}$)  
will allow a more careful evaluation of the shadowing corrections at 
small $x$ \cite{prd,ayala3}.

\section{ The radius dependence of the shadowing corrections}

The value of SC crucially depends on the size of the target \cite{hotspot}. In pQCD 
the value of $R$ is associated with the coupling of  the gluon ladders with the 
proton, or to put it in  another way, on how the gluons are distributed within
the proton. $R$ may be of  the order of the proton radius if the gluons are distributed 
uniformly in the whole proton disc or much smaller
if the gluons are concentrated, {\it i.e.} if the gluons in the proton are confined in a 
disc with smaller radius than the size of the proton.

 Considering the expression (\ref{f2eik}),  assuming $\Omega_{q\overline{q}} < 1$ 
and  expanding the expression to ${\cal{O}}(\Omega^2)$ we obtain
\begin{eqnarray}
F_2(x,Q^2) =  \frac{1}{2\pi^3} \sum_{u,d,s} e_f^2 \int_{\frac{1}{Q^2}}^{\frac{1}{Q_0^2}} \frac{d^2r_t}{r_t^4} \int d^2b_t 
\left\{\frac{1}{2}\Omega_{q\overline{q}} - \frac{1}{8}\Omega^2_{q\overline{q}}\right\}\,\,.
\label{f2eikexp}
\end{eqnarray}
Using the factorization of the opacity and the normalization of the profile function we 
can write  $F_2$ as
\begin{eqnarray}
F_2(x,Q^2) =  \frac{1}{2\pi^3} \sum_{u,d,s} e_f^2 \int_{\frac{1}{Q^2}}^{\frac{1}{Q_0^2}} \frac{d^2r_t}{r_t^4}  
\left\{\frac{1}{2}\overline{\Omega_{q\overline{q}}} - \frac{1}{8}\overline
{\Omega_{q\overline{q}}}^2\int d^2b_t S^2(b_t)\right\}\,\,.
\label{f2eikexp2}
\end{eqnarray}
The second term of the above equation represents the first shadowing corrections for 
the $F_2$ structure function. Assuming a Gaussian parametrization 
for the profile function we obtain that  the screening is inversely proportional to the radius. 
Therefore the shadowing corrections are strongly associated with the distributions of the 
gluons within  the proton. In this section we estimate the radius dependence of the 
shadowing corrections, considering the  $F_2$ and $\frac{dF_2(x,Q^2)}{dlogQ^2}$ 
data. First we explain why the radius is  expected to be smaller than the proton radius. 

Consider the first order contribution to the shadowing corrections, where two  ladders 
couple to the proton.  The  ladders may be attached to  different constituents of the 
proton or to the same constituent. In the first case the shadowing corrections are 
controlled by the proton radius, while in the second case these corrections are controlled
by the constituent radius, which is  smaller than the proton radius.   Therefore, on the 
average, we expect that the radius will be smaller than the proton radius. Theoretically, 
 $R^2$ 
reflects the integration over $b_t$ in the first diagrams for the SC.

\begin{figure}
\centerline{\psfig{file=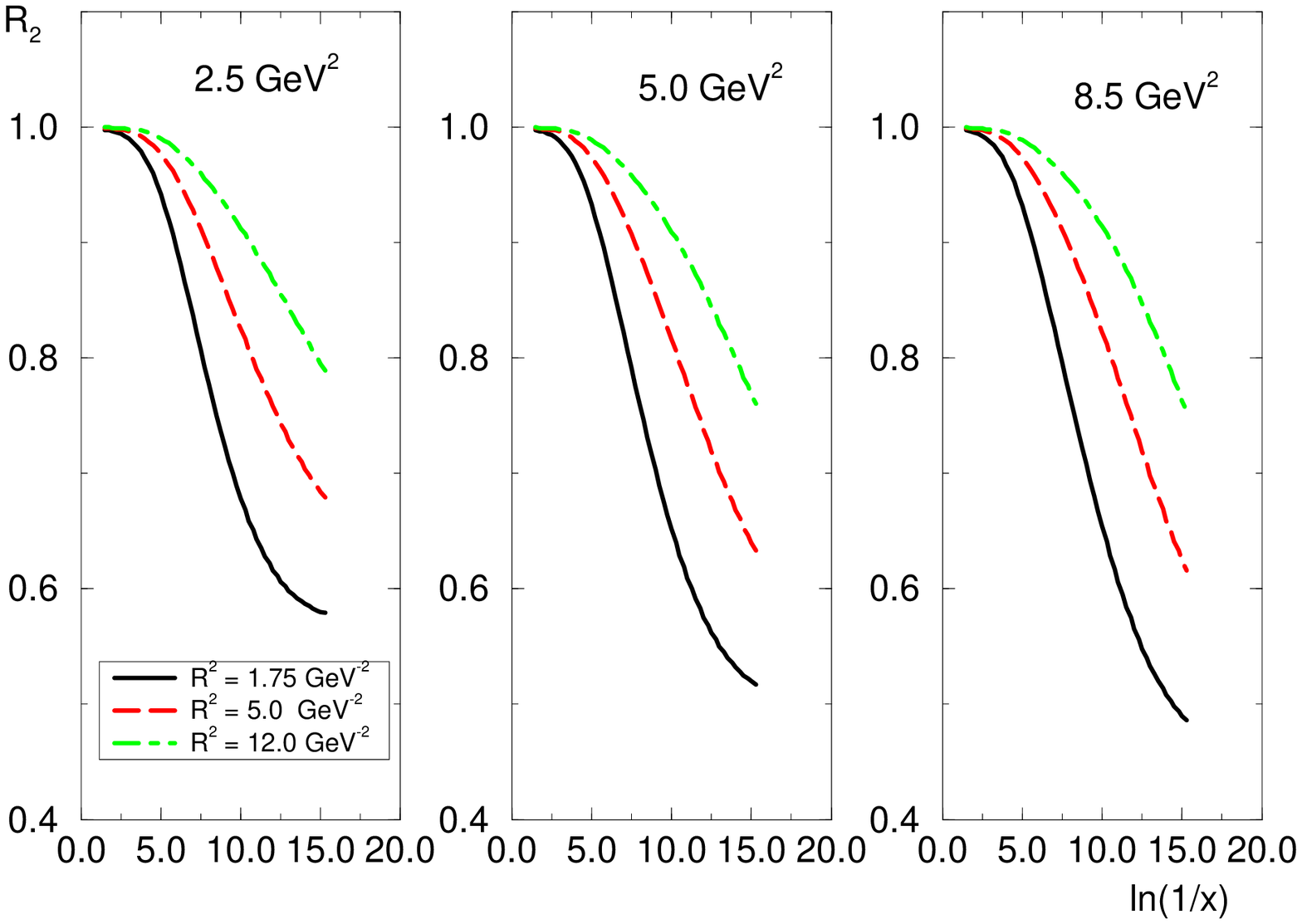,width=150mm}}
\caption{ The ratio $R_2 = \frac{F_2(x,Q^2)\mbox{[Eq. 
(\ref{diseik2})]}}{F_2(x,Q^2)\mbox{[GRV]}}$ as a  function of the 
variable $ln(\frac{1}{x})$ for different virtualities and radii.}
\label{fig3}
\end{figure}

\begin{figure}
\centerline{\psfig{file=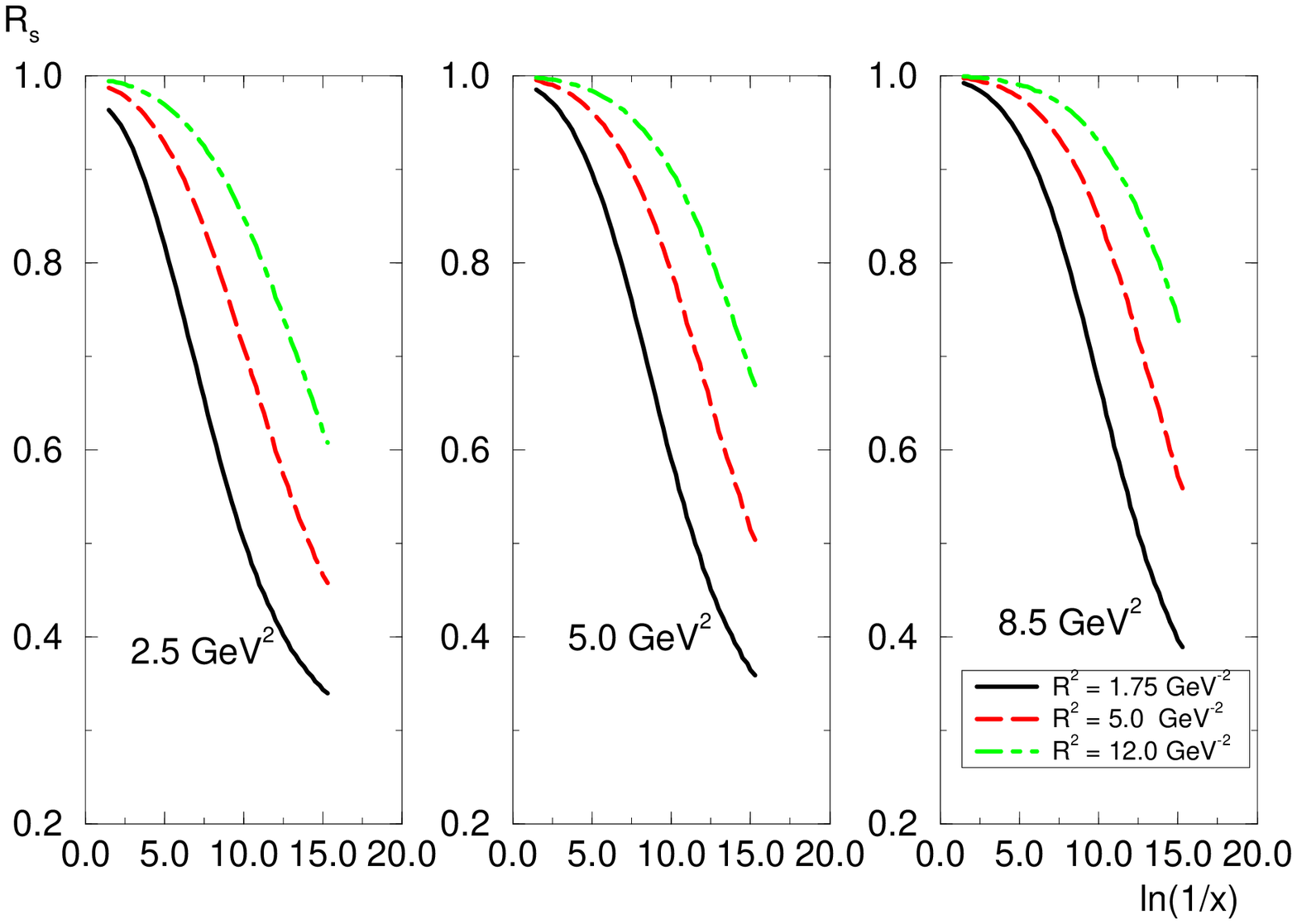,width=150mm}}
\caption{ The ratio $R_s = \frac{\frac{dF_2(x,Q^2)}{dlogQ^2}\mbox{[Eq. (\ref{df2eik})]}}{\frac{dF_2(x,Q^2)}{dlogQ^2}\mbox{[GRV]}}$   as a  function of the variable $ln(\frac{1}{x})$ for different virtualities and radii.}
\label{fig4}
\end{figure}

 In Fig. \ref{fig3}
we present the ratio
\begin{eqnarray}
R_2 = \frac{F_2(x,Q^2)\mbox{[Eq. (\ref{f2})]}}{F_2(x,Q^2)\mbox{[GRV]}}\,\,,
\label{r2}
\end{eqnarray}
where 
$F_2(x,Q^2) [\mbox{GRV}] = \sum_{u,d,s} e_f^2 \,[xq(x,Q^2) + x
\overline{q}(x,Q^2)] + F_2^c(x,Q^2)$
is calculated using the GRV parametrization.
For the treatment of the charm component of the structure function we consider the charm 
production via boson-gluon fusion \cite{grv95}. In this paper we assume $m_c = 1.5\,GeV$. 
In Fig. \ref{fig4} we present the ratio
\begin{eqnarray}
R_s = \frac{\frac{dF_2(x,Q^2)}{dlogQ^2}\mbox{[Eq. (\ref{df2})]}}{\frac{dF_2(x,Q^2)}{dlogQ^2}\mbox{[GRV]}}\,\,.
\label{rs}
\end{eqnarray}
The function $\frac{dF_2(x,Q^2)}{dlogQ^2}\mbox{[GRV]}$ was calculated using the 
expression (\ref{df2glap}) and the GRV parametrization.
Our results are presented as a function of $ln(\frac{1}{x})$ at different virtualities. 
We can see that the  SC are larger in the ratio $R_s$ and that our predictions of SC 
are strongly dependent of the radius $R$. Moreover, we see clearly the SC behavior 
inversely proportional  with the radius. 

\begin{figure}
\centerline{\psfig{file=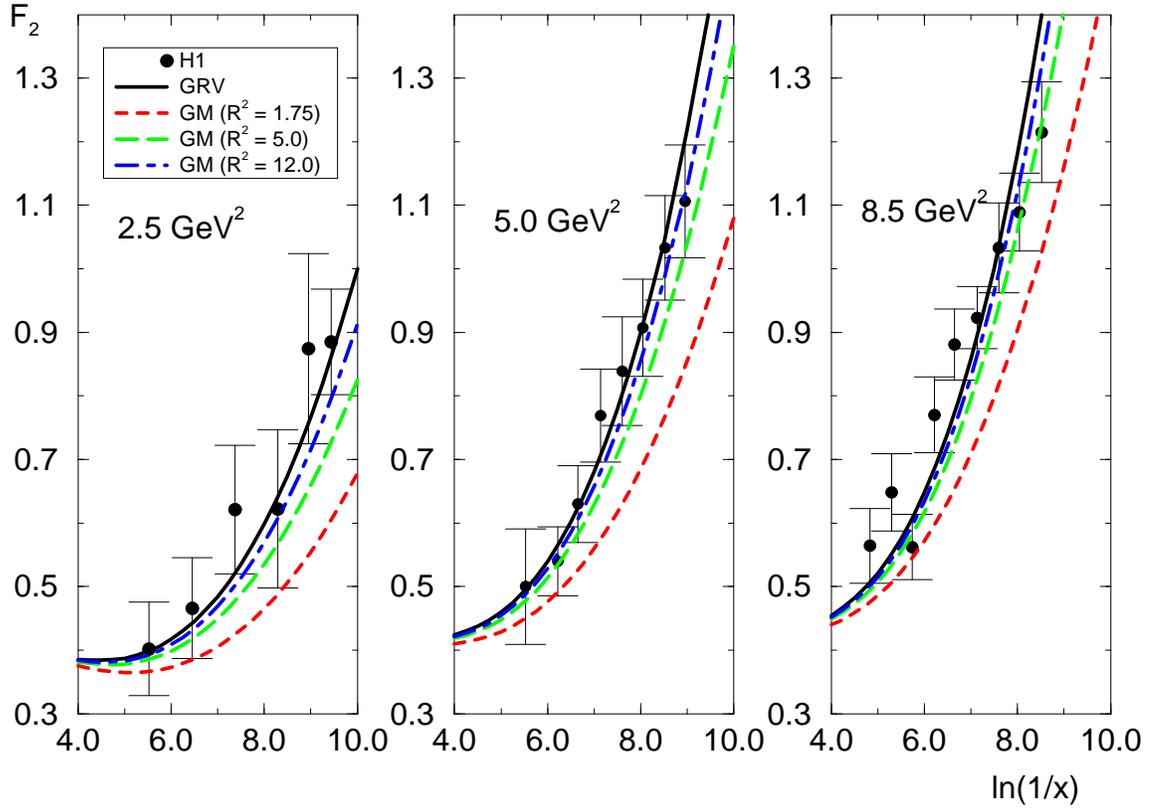,width=150mm}}
\caption{ The $F_2$ structure function as a function of the variable $ln(\frac{1}{x})$ 
for different virtualities and radii.  Only the shadowing corrections in the quark sector 
are considered. Data from H1 \cite{h1}.}
\label{fig5}
\end{figure}

 In Fig. \ref{fig5} we compare our predictions for the SC in the $F_2$ structure 
function and the H1 data \cite{h1} as a function of $ln(\frac{1}{x})$ at different 
virtualities and some values of the radius. Our goal is not a best fit of the radius, but 
eliminate some values of radius  comparing the predictions of the eikonal approach 
and  HERA data.  We  consider only the quark sector in the calculation of SC, which  
is a good approximation in this observable,  as shown in the previous section. The 
choice $R^2 = 1.5 \, GeV^2$  does not describe the data, {\it i.e.} the data discard 
the possibility of  very large SC in the HERA kinematic region. However, there are still 
two possibilities for the radius which reasonably describe the  $F_2$ data.  To discriminate 
between these possibilities we must consider the behavior of the  $F_2$ slope. 

\begin{figure}
\centerline{\psfig{file=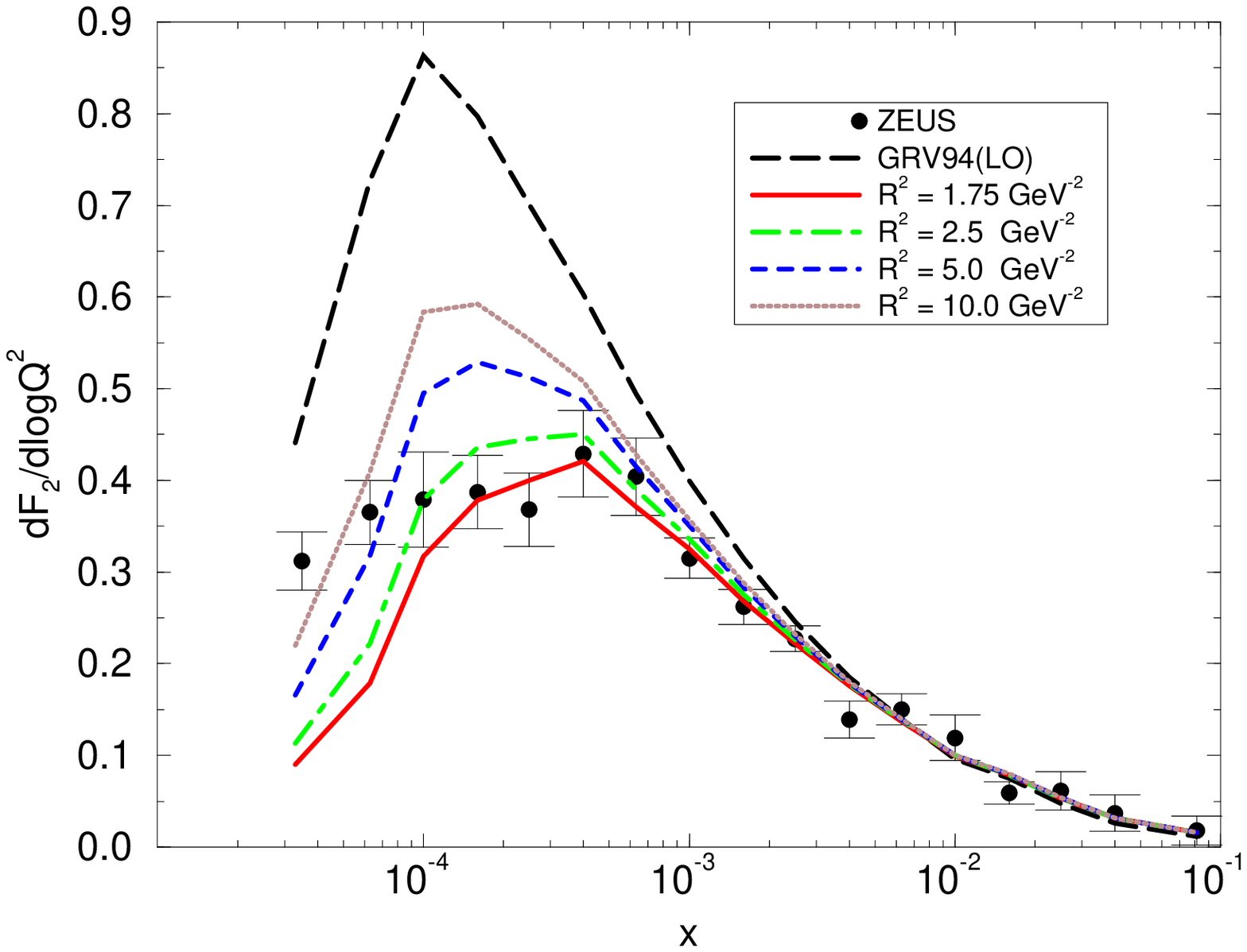,width=150mm}}
\caption{The $F_2$  slope  as a function of the variable $x$ for different radii. 
Only the shadowing corrections in the quark sector are considered.  Data from 
ZEUS \cite{zeus}. The data points correspond to a different $x$ and $Q^2$ value.}
\label{fig6}
\end{figure}

\begin{figure}
\centerline{\psfig{file=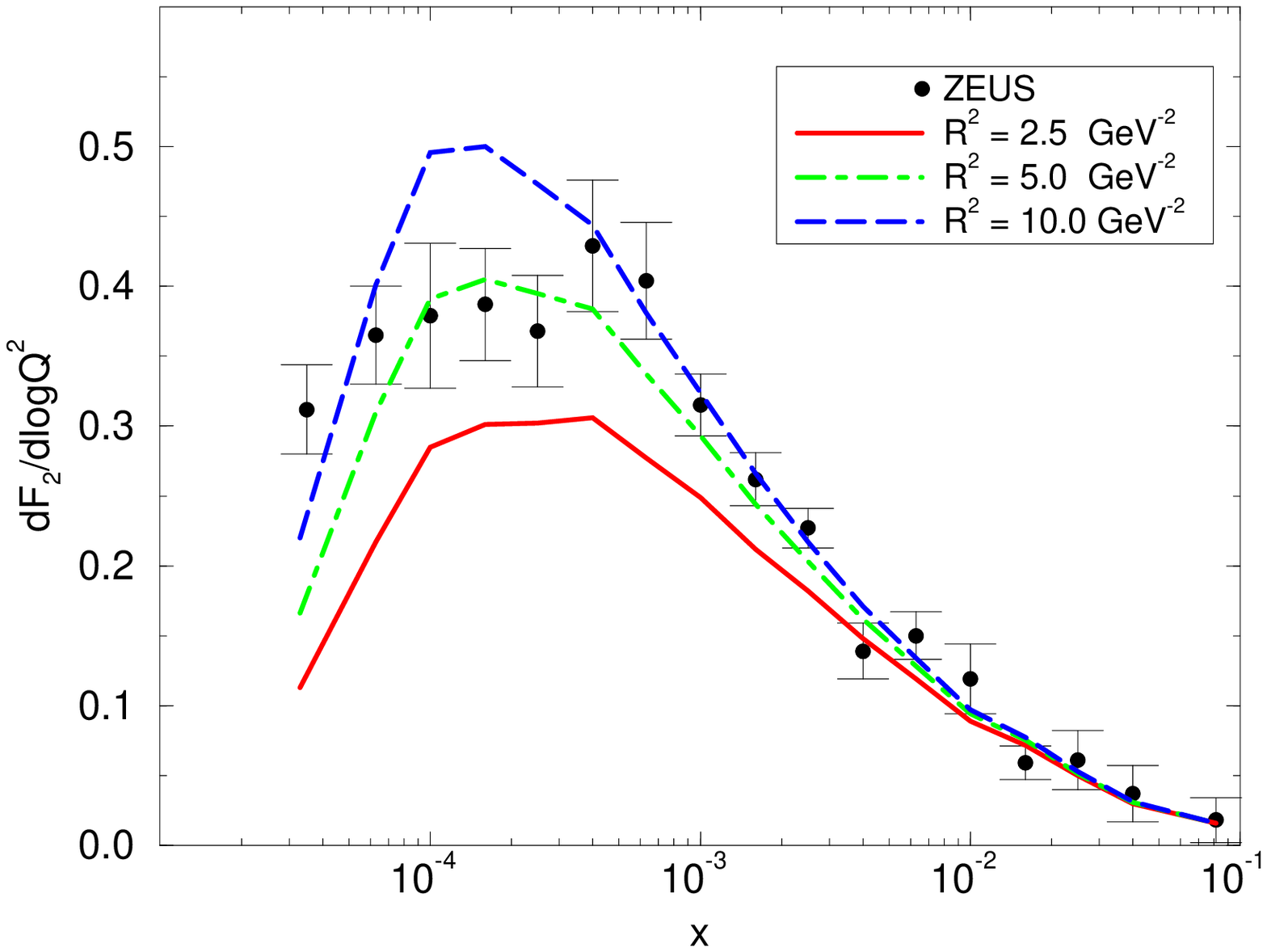,width=150mm}}
\caption{The $F_2$  slope  as a function of the variable $x$ for different radii.  
The shadowing corrections in the gluon-quark sector are considered.  Data from 
ZEUS \cite{zeus}. The data points correspond to a different $x$ and $Q^2$ value.}
\label{fig7}
\end{figure}

In Fig. \ref{fig6} we present our results for  $\frac{dF_2(x,Q^2)}{dlogQ^2}$ considering 
the SC only in the quark sector. Although in the previous section we have demonstrate that 
the contributions of the quark and gluon sectors should be considered,
 here we will test other possibilities to describe the data: the dependence on the radius $R$. 
Our results show that the best fit of the data occurs at small values of $R^2$,  which  are 
discarded by the $F_2$ data. Therefore, in agreement with our previous conclusions, we 
must consider a general approach to describe consistently the $F_2$ and 
$\frac{dF_2(x,Q^2)}{dlogQ^2}$ data. In Fig. \ref{fig7} we present our results for  $\frac{dF_2(x,Q^2)}{dlogQ^2}$ considering the SC in the gluon and quark sector for 
different values of $R^2$, calculated using the general approach proposed in the previous section.
The best result occurs for $R^2 = 5\,GeV^{-2}$, which also describes the $F_2$ data.

The value for the squared radius $R^2 = 5\,GeV^{-2}$ obtained in our analysis agrees 
with the estimates obtained using the  
HERA data on diffractive photoproduction of $J/\Psi$ meson \cite{zeusjpsi,h1jpsi}. 
Indeed, the experimental values for the slope are
$B_{el} = 4 \, GeV^{-2}$ and $B_{in} = 1.66\,GeV^{-2}$ and the cross section
for $J/\Psi$ diffractive production with and without photon dissociation
are equal. Neglecting the $t$ dependence of the pomeron-vector meson
coupling  the value of $R^2$ can be estimated \cite{plb}. It turns out that 
$R^2 \approx 5\,GeV^{-2}$, {\it i.e.}, approximately 2 times smaller than 
the radius of the proton.

As an 
additional comment let us say that the SC to $F_2$ and 
its slope may also be analysed using a two radii model for the 
proton \cite{glmn1}. This analysis is motivated by the large difference 
between the measured slopes in elastic and inelastic diffractive 
leptoproduction of vector mesons in DIS. An analysis using the   two radii 
model for the proton is not a goal  of this paper, since a definite 
conclusion on the correct model is still under debate. 

The  summary of this point is that the analysis of the $F_2$ and $\frac{dF_2(x,Q^2)}
{dlogQ^2}$ data using the eikonal model implies that the gluons are not distributed 
uniformly in the whole proton disc, but behave as concentrated in smaller regions. 
This conclusion motivates an analysis of the jet production, which probes smaller 
regions within the proton, using an approach which considers the shadowing corrections.

\section{A screnning boundary}

The common feature of the BFKL and DGLAP equations  is the steep increase  of the 
cross sections as $x$ decreases. This steep increase cannot persist down to arbitrary 
low values of $x$ since it violates a fundamental principle of quantum theory, {\it i.e.} the 
unitarity. In the context of relativistic quantum  field theory of the strong interactions, unitarity 
implies  the cross section of a hadronic scattering reaction cannot increase with increasing 
energy $s$ above $log^2 \,s$: the Froissart's theorem \cite{froi}. The Froissart 
bound cannot be proven for off-mass-shell amplitudes \cite{yndu}, which is the case for 
deep inelastic scattering \cite{glmuni}. 
Our goal in this section is by using the $s$-channel unitarity (\ref{uni}) and the eikonal approach 
to estimate a superior limit from which the shadowing corrections cannot be disregarded in $F_2$ 
and its slope.

 Considering the   expression (\ref{f2eik}) for the $F_2$ structure function, 
we can write a  $b_t$ dependent structure function given by
\begin{eqnarray}
F_2(x,Q^2,b_t) =  \frac{1}{2\pi^3} \sum_{u,d,s} e_f^2 \int_{\frac{1}{Q^2}}^{\frac{1}{Q_0^2}} \frac{dr^2_t}{r_t^4}  
\left\{1 - e^{-\frac{1}{2}\Omega_{q\overline{q}}(x,r_t,b_t)}\right\}\,\,.
\label{f2eikuni}
\end{eqnarray}

The relation between the opacity and the gluon distribution (\ref{omega}) obtained in 
\cite{plb}, is valid in the kinematical region where $\Omega \ll 1$. In the eikonal 
approach for pQCD we make the assumption that the relation (\ref{omega}) is valid in 
all kinematic region. 
To obtain an estimate of the region where the SC are important we consider  
a superior limit for the expression (\ref{f2eikuni}), which occurs for $\Omega \gg 1$. In this limit
the second term in the above equation can be disregarded. As the shadowing 
terms are negative and reduce the growth of  the $F_2$ structure function, 
disregarding the shadowing terms we are estimating a superior limit for the 
region where these terms are not important, {\it i.e.} a screnning boundary 
which establishes the region where the shadowing corrections are required to calculate the observables.

The  $b_t$ dependent  structure function in the limit  $\Omega \gg 1$ is such that
\begin{eqnarray}
F_2(x,Q^2,b_t) <  \frac{1}{2\pi^3} \sum_{u,d,s} e_f^2 \int_{\frac{1}
{Q^2}}^{\frac{1}{Q_0^2}} \frac{dr^2_t}{r_t^4}  \,\,.
\label{f2eikuni3}
\end{eqnarray}
Making the assumption that  the $b_t$ dependence of the structure function is factorized \cite{plb}: 
\begin{eqnarray}
 F_2(x,Q^2,b_t)  =  F_2(x,Q^2) \, S(b_t) \nonumber \,\,,
\end{eqnarray} 
and considering a Gaussian parametrization for the profile function and its value 
for $b_t = 0$ we get ($n_f = 3$)
\begin{eqnarray}
F_2(x,Q^2) <  \frac{R^2}{3\pi^2}  \int_{\frac{1}{Q^2}}^{\frac{1}{Q_0^2}}
 \frac{dr^2_t}{r_t^4}  \,\,.
\label{f2eikuni4}
\end{eqnarray}
As a result
 \begin{eqnarray}
F_2(x,Q^2) & <  & \frac{R^2}{3\pi^2} (Q^2 - Q_0^2) \nonumber \\
& < & \frac{R^2 Q^2}{3\pi^2} \,\,.
\label{unif2}
\end{eqnarray}
The above limit is our estimate for the screnning boundary for the $F_2$ structure function.

The screnning boundary for the $F_2$ slope is straightforward from the   
 expression (\ref{f2eikuni3}).  We  get 
\begin{eqnarray}
\frac{dF_2(x,Q^2)}{dlogQ^2} <  \frac{R^2 Q^2}{3\pi^2} \,\,. 
\label{unidf2}
\end{eqnarray}
This expression agrees with the expression obtained in \cite{plb}. 

Clearly expressions  (\ref{unif2}) and  (\ref{unidf2}) serve only as a rough prescription 
for estimating the region where the corrections required by unitarity cannot be disregarded. A more 
rigorous treatment would be desirable, but remains to be developed.  

\begin{figure}
\centerline{\psfig{file=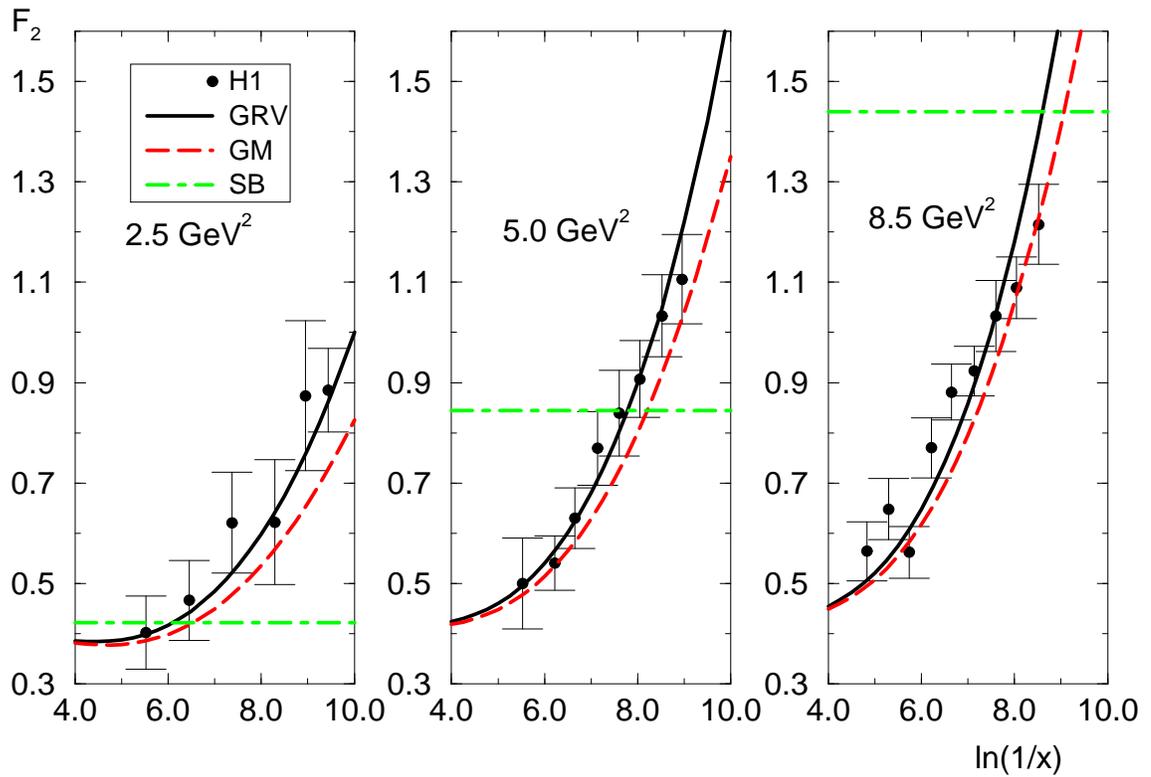,width=150mm}}
\caption{ Screnning boundary to the $F_2$ structure function.  For points above  the boundary the  corrections cannot be disregarded. Data from H1 \cite{h1}.}
\label{fig8}
\end{figure}

\begin{figure}
\centerline{\psfig{file=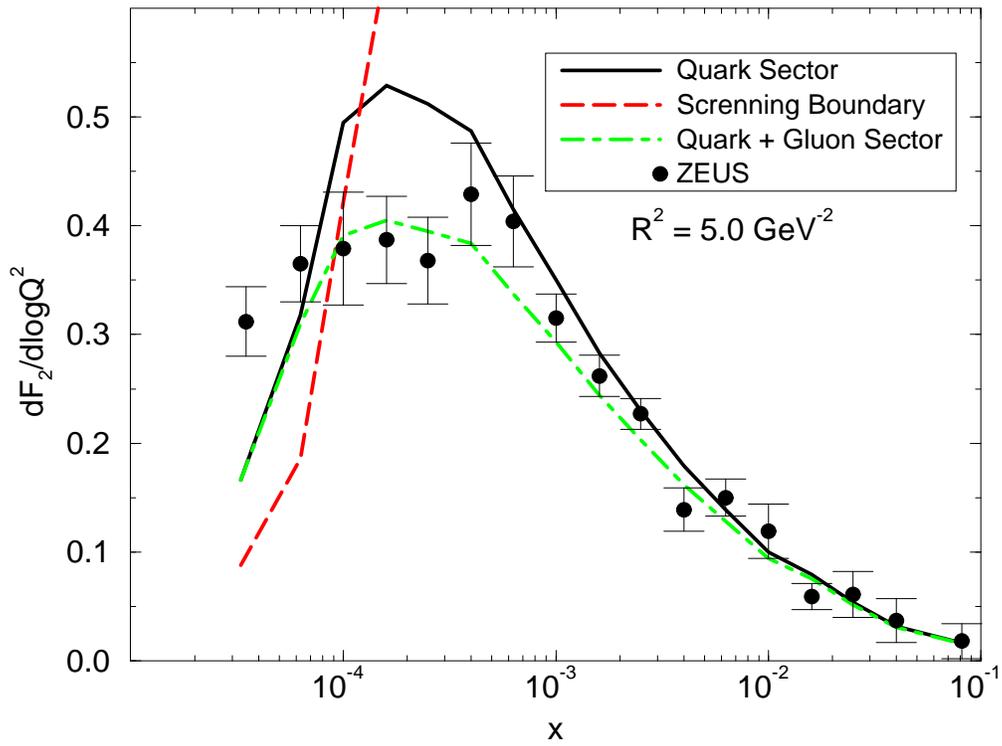,width=150mm}}
\caption{ Screnning boundary to the $F_2$ slope.  For points above  the boundary the  corrections cannot be disregarded. Data from  ZEUS \cite{zeus}.}
\label{fig9}
\end{figure}

Using the above expressions we can make an analysis of  HERA data. We use $R^2 = 5 
\, GeV^{-2}$ in the calculations. In Fig. \ref{fig8} we compare our predictions   
with the $F_2$ data from the H1 collaboration. We can see that data at larger values of 
$Q^2$ ($Q^2 \ge 8.5 \, GeV^2$) do not violate the limit (\ref{unif2}). However, the 
data at smaller values of $Q^2$ and $x$  violate this limit. This indicates 
that we should consider the SC for this kinematical region. In Fig. \ref{fig9} we present 
our results for the $F_2$ slope. We see that the data  for small $Q^2$ 
and $x$ ($Q^2 \le 2.5 \,GeV^2$, $x \le 10^{-4}$) violate 
the limit  (\ref{unidf2}), stressing the need of the shadowing corrections. 
Therefore for small values of $x$ and $Q^2$  the observables  must be calculated using 
an approach which takes them into account.

\section{Summary}

In this paper we have presented our analysis of the shadowing corrections in the scaling 
violations using the eikonal approach.  We shown that the $\frac{dF_2(x,Q^2)}{dlogQ^2}$ 
data can be described successfully considering the shadowing corrections in the quark and gluon sectors.  
 Furthermore, we have considered the radius dependence of these corrections and  an 
unitarity boundary. From the analysis of the $R$ dependence of the SC  in the eikonal 
approach we have shown that   the value $R^2 = 5\, GeV^{-2}$ allows to describe 
the HERA data. This value agrees with the estimate obtained independently in the 
diffractive $J/\Psi$ photoproduction. Using the eikonal approach and the assumption of $b_t$ factorization 
 of the $F_2$ structure function a screnning boundary is analysed.
 This boundary constrains the region where the  corrections required by unitarity may be disregarded;
or in other words, a limit for applicablity of standard perturbative QCD framework. 
We have shown that the HERA data at small $x$ and $Q^2$
violate this limit, which implies that the shadowing corrections are important 
in the HERA kinematic region.

Our conclusion is that the shadowing effect is important already at HERA kinematic region.
We believe that   the  analysis of distinct observables  
 ($F_L, \, F_2^c, \, \frac{dF_2(x,Q^2)}{dlogQ^2}$) at small values of $x$ 
 and $Q^2$ will allow to evidentiate the shadowing corrections.

\section*{Acknowledgments}

MBGD acknowledges enlightening discussions with F. Halzen at University of Wisconsin, 
S. J. Brodsky at SLAC  and E. M. Levin during the completion of this work.

\newpage

\end{document}